\documentclass{article}
\usepackage{spconf,amsmath,graphicx}

\usepackage{amsfonts}
\usepackage{multirow}
\usepackage{hyperref}
\usepackage{subfigure}
\usepackage{booktabs,setspace}
\usepackage{threeparttable}
\ninept
\usepackage{setspace}

\usepackage{amssymb,CJK, indentfirst,balance,appendix}

\title{Hourglass-AVSR: Down-Up Sampling-based Computational Efficiency Model for Audio-Visual Speech Recognition}
\name{\begin{tabular}{c}Fan Yu, Haoxu Wang, Ziyang Ma, Shiliang Zhang\end{tabular}}

\address{Speech Lab of DAMO Academy, Alibaba Group, China
        }
\begin{document}

%\ninept
%
\maketitle

\begin{abstract}
Recently audio-visual speech recognition (AVSR), which better leverages video modality as additional information to extend automatic speech recognition (ASR), has shown promising results in complex acoustic environments.
However, there is still substantial space to improve as complex computation of visual modules and ineffective fusion of audio-visual modalities.
To eliminate these drawbacks, we propose a down-up sampling-based AVSR model (Hourglass-AVSR) to enjoy high efficiency and performance, whose time length is scaled during the intermediate processing, resembling an hourglass.
Firstly, we propose a context and residual aware video upsampling approach to improve the recognition performance, which utilizes contextual information from visual representations and captures residual information between adjacent video frames.
Secondly, we introduce a visual-audio alignment approach during the upsampling by explicitly incorporating boundary constraint loss. 
Besides, we propose a cross-layer attention fusion to capture the modality dependencies within each visual encoder layer.
Experiments conducted on the MISP-AVSR dataset reveal that our proposed Hourglass-AVSR model outperforms ASR model by 12.9\% and 20.8\% relative concatenated minimum permutation character error rate (cpCER) reduction on far-field and middle-field test sets, respectively.
%, which achieves the highest cpCER improvement of visual module compared with other SOTA AVSR models.
Moreover, compared to other state-of-the-art AVSR models, our model exhibits the highest improvement in cpCER for the visual module.
Furthermore, on the benefit of our down-up sampling approach, Hourglass-AVSR model reduces 54.2\% overall computation costs with minor performance degradation.

\end{abstract}

\begin{keywords}
Multi-Modal Processing, Audio-Visual Speech Recognition, Computational Efficiency, Down-Up Sampling
\end{keywords}
\vspace{-5pt}
\section{Introduction}
\label{sec:intro}
\vspace{-3pt}
Recent years have witnessed the explosion of automatic speech recognition (ASR), with the advances in deep learning and end-to-end (E2E) neural approaches~\cite{li2021recent}.
%In recent years, with the advances in deep learning, end-to-end neural approaches have rapidly gained prominence in the automatic speech recognition (ASR) community~\cite{li2021recent}.
However, ASR is still a challenging problem in complex acoustic conditions, such as overlapping speech, noise, reverberation, etc~\cite{Yu2022M2MeT,Yu2022Summary,fiscus2007rich,fiscus2006rich, chen2022audio}.
Fortunately, numerous researches have demonstrated that visual information can provide effective cues to aid speech recognition, especially in such complex acoustic environment where the audio is corrupted or unavailable.
Therefore, audio-visual speech recognition (AVSR), incorporating video modality (e.g. lip motion and facial expression) into ASR, has been extensively developed~\cite{tamura2015audio,petridis2018end,makino2019recurrent}.

Current works on AVSR major focus on improving the model performance through optimizations in model architecture and modality fusion.
Several well-established architectures in ASR have been adapted for AVSR model, such as Element-wise-Attention Gated Recurrent Unit (EleAtt-GRU)~\cite{xu2020discriminative}, Recurrent Neural Network Transducer (RNN-T)~\cite{makino2019recurrent}, Watch, Listen, Attend and Spell (WLAS)~\cite{son2017lip}, Transformer~\cite{deep2022afouras} and Conformer~\cite{ma2021end}, etc.
In terms of modality fusion, previous approaches commonly employed direct concatenation of audio and visual features as the model input~\cite{xu2023nio,xu2022channel}. 
Although these approaches reduce the parameters without a visual front-end and encoder, they often struggles to effectively capture the representations of the visual modality.
%To achieve better performance, audio-visual bi-encoder framework was proposed to learn each modality representations independently,  and adopted multi-layer perception 
To achieve better performance, Ma et al.~\cite{ma2021end} adopted a bi-encoder framework to learn each modality representations and multi-layer perception to fuse modality information.
%which uses ResNet as the visual encoder and Conformer as the audio encoder, and multi-layer perception (MLP) to fuse modality information.
With a similar audio-visual bi-encoder framework, cross-attention fusion was also proposed to combine the audio and visual representations from different modality encoders~\cite{sterpu2018attention, wei2020attentive,wu2021audio}, which established temporal alignments between audio and video sequences.
Additionally, cross-layer fusion methods have been introduced to capture multi-level modality knowledge by summing the hidden representations within each audio and visual encoder layer~\cite{li2023xmu}.

Computational efficiency is also an essential aspect for AVSR research, but it is often overlooked in favor of performance improvements.
Many E2E AVSR models employ an audio-visual bi-encoder framework that requires a strong visual front-end to encode visual streams, such as lip motion, into high-level representations for recognition.
Typically, the visual front-end is a trainable 3-dimensional convolutional network (3D ConvNet), such as 3D version of ResNet~\cite{he2016deep} or VGG~\cite{simonyan2014very}.
Despite its good performance, 3D ConvNet is computationally intensive and becomes intractable on AVSR models, which brings larger computation costs compared with the audio front-end~\cite{tran2015learning}.
There are works in video understanding that have been attempted to trade off expressiveness and computation costs~\cite{lee2018motion,lin2019tsm,tran2018closer,xie2018rethinking,shou2017cdc}.
To save computation, motion filter~\cite{lee2018motion} and temporal shift module (TSM)~\cite{lin2019tsm} were proposed to generate spatio-temporal features from 2D ConvNet.
Meanwhile, Tran et al.~\cite{tran2018closer} and Xie et al.~\cite{xie2018rethinking} introduced mixed 2D-3D convolutional network to achieve computational efficiency in AVSR models.
In addition to the structural optimization, Shou et al.~\cite{shou2017cdc} proposed a convolutional-de-convolutional filter (CDC) that performs convolution-based sampling to reduce computation cost.

In this paper, we propose a down-up sampling-based computational efficiency model for AVSR (Hourglass-AVSR) that enjoys both high efficiency and high performance.
Considering that AVSR is a multi-modal fusion model which may utilize different visual modules in various scenarios, we aim to develop a generalized and robust approach for reducing computing costs across different visual modules.
Down-up sampling seems an efficient solution, typically achieved through convolution and de-convolution.
However, the de-convolution upsampling approach primarily focuses on leveraging context information while overlooking the lost information during downsampling.
To address this limitation, we propose a context and residual aware video upsampling approach to compensate for the performance loss caused by the downsampling, which consists of contextual-based prediction and residual-based prediction modules.
The contextual-based prediction module captures the temporal relationship of visual representations and extracts fine-grained context-wise patterns.
While the residual-based prediction module utilizes the video sequence before downsampling to calculate residuals information, which conveys the movement of speaker mouth.
Note that the residual information can estimate motion implicitly and improve the precision of visual representation prediction.
%The residual information between adjacent video frames conveys the movement of speaker mouth, which estimates motion implicitly and improves the precision of visual representation prediction.
%By integrating these two modules, our approach effectively addresses the limitations of the traditional de-convolution upsampling approach. It not only leverages contextual information but also incorporates residual information to enhance the accuracy and quality of visual representations during upsampling.

Compared with the video-only upsampling, the upsampling of AVSR can use audio representations as additional information.
Inspired by the temporal alignment between video and audio, we introduce a boundary constraint loss to explicitly model the visual-audio alignment during upsampling, which further improves the prediction accuracy.
Furthermore, we propose a cross-layer attention fusion to capture modality dependencies within each encoder layer.
Different from previous cross-layer fusion~\cite{li2023xmu}, our cross-layer attention fusion employs cross-attention to achieve more fine-grained modeling of audio-visual representations, rather than simple summation.
Experiments on the MISP-AVSR dataset show that our proposed Hourglass-AVSR model outperforms the audio-only ASR model by 12.9\% and 20.8\% relative concatenated minimum permutation character error rate (cpCER) reduction on far-field and middle-field test sets, which achieves the highest cpCER improvement of the visual module compared with other SOTA AVSR models.
Moreover, on the benefit of our sampling approach, our model reduces 54.2\% overall computation costs with minor performance degradation.

\vspace{-10pt}
\section{PROPOSED METHOD}
\label{sec:method}
\vspace{-9pt}
\subsection{Overview}
\begin{figure}[!htb]
	\centering
	\includegraphics[scale=0.54]{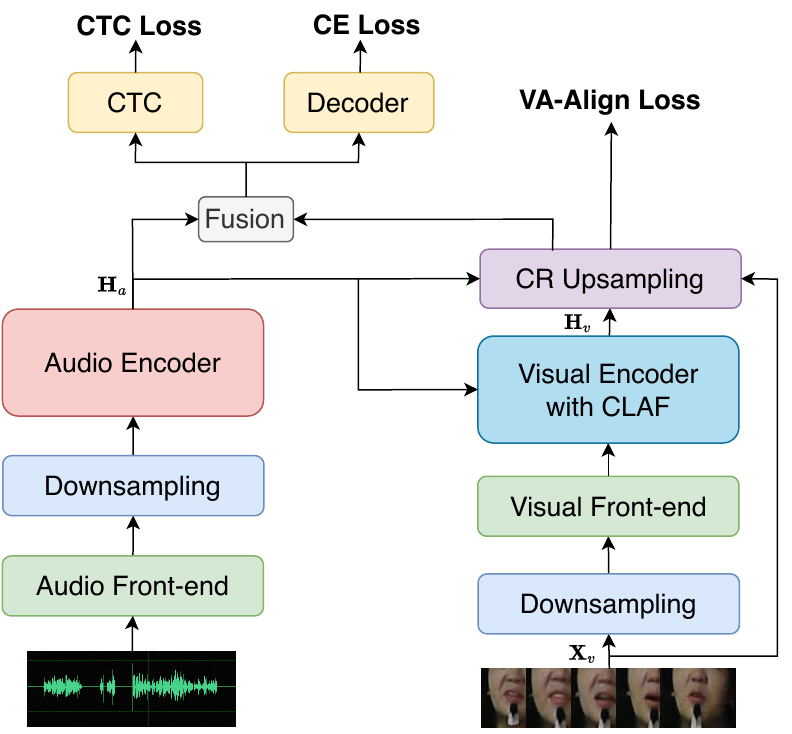}
        \vspace{-5pt}
        \caption{
		An overview of the Hourglass-AVSR model. \textit{CR Upsampling} means context and residual aware video upsampling module; \textit{VA-Align Loss} means visual-audio alignment loss; \textit{Visual Encoder with CLAF} means visual encoder with cross-layer attention fusion.
	}
        \label{model_all}
 \vspace{-18pt}
\end{figure}

\vspace{-4pt}
The overall framework of our proposed Hourglass-AVSR model is illustrated in Fig. \ref{model_all}, which adopts audio-visual bi-encoder architecture.
Hourglass-AVSR model consists of eight modules, namely the downsampling, audio front-end, visual front-end, audio encoder, visual encoder with cross-layer attention fusion (CLAF), context and residual aware (CR) video upsampling, decoder and loss function.
In detail, audio data is first processed through the front-end module for feature extraction and later downsampling, while video data is processed in reverse order due to the large computation costs of the visual front-end.
The audio front-end extracts mel-filterbank features from the audio signals, while the visual front-end takes RGB mouth tracks as the input.
Both the audio and visual encoders consist of multiple blocks of Branchformer~\cite{peng2022branchformer} for learning modality-specific knowledge.
Especially, cross-layer attention fusion is introduced to the visual encoder for modeling the relationship between audio and visual at each layer.
Then, the context and residual aware video upsampling module restores the original length of video sequence, and visual-audio alignment loss is introduced during the upsampling.
Finally, the representations of different modalities are fused to calculate connectionist temporal classification (CTC) loss, and the fused representations are fed into decoder for recognition and cross-entropy (CE) loss calculation.

\vspace{-8pt}
\subsection{Context and Residual Aware Upsampling}
\vspace{-2pt}

\begin{figure*}[t]
	\centering
 \vspace{-5pt}
	\includegraphics[width=0.85\linewidth]{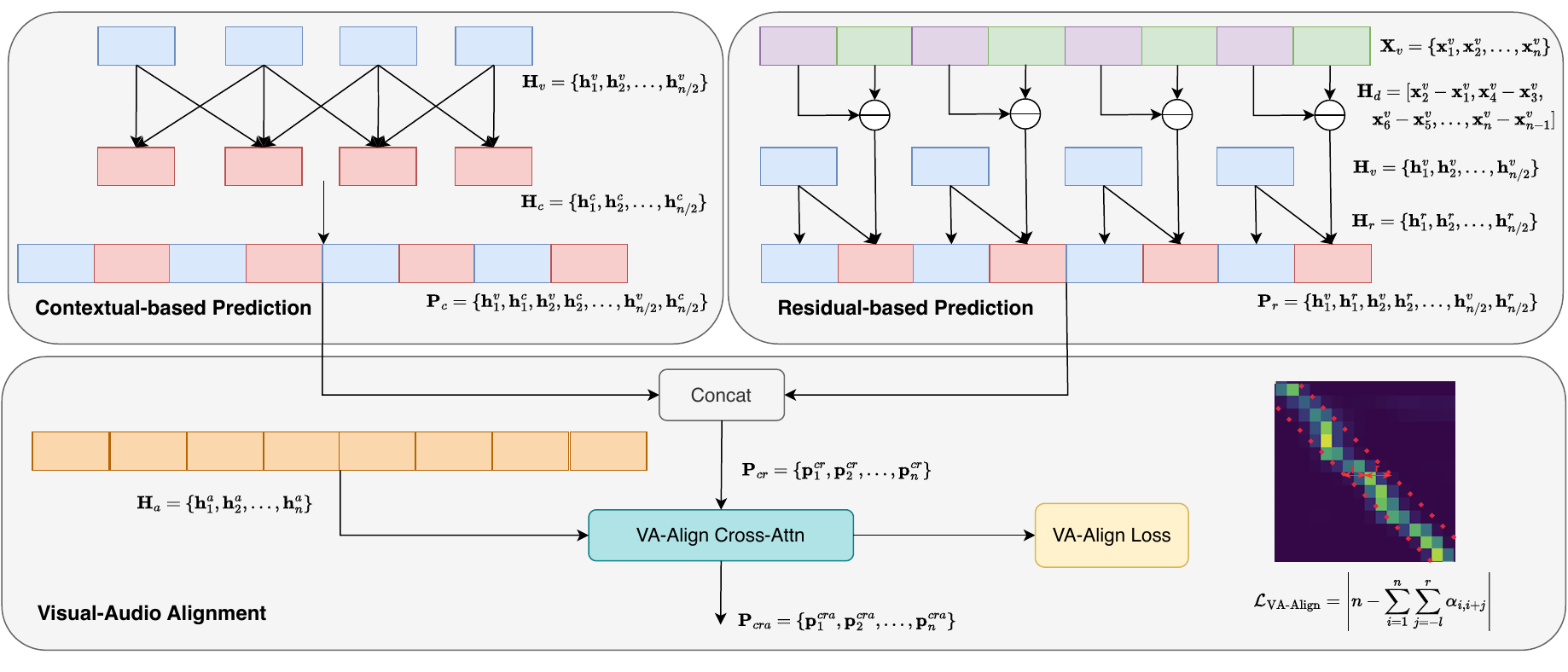}
  \vspace{-13pt}
	\caption{
	       The architecture of the proposed context and residual aware upsampling (upper) and visual-audio alignment loss (lower).
	}

	\label{upsample}
 \vspace{-17pt}
\end{figure*}

Context and residual aware video upsampling module consists of contextual-based prediction module and residual-based prediction module, as show in Fig \ref{upsample}.
The visual encoder output $\mathbf{H}_v=\{\mathbf{h}_1^v,\mathbf{h}_2^v,...,\mathbf{h}_{n/d}^v\}$, audio encoder output $\mathbf{H}_a=\{\mathbf{h}_1^a,\mathbf{h}_2^a,...,\mathbf{h}_{n}^a\}$ and video sequence before downsampling $\mathbf{X}_v=\{\mathbf{x}_1^v,\mathbf{x}_2^v,...,\mathbf{x}_n^v\}$ are as the input fed into the upsampling module, where $n$ is the length and $d$ is the factor of downsampling.
In order to clearly show the whole upsampling process, the figure uses a $2\times$ upsampling ($d=2$) as an example.
Noting that we control the length of the audio sequence through the audio frontend and downsampling, keeping the same length as the video sequence.
The contextual-based prediction module captures the temporal context of the visual representations for predicting, which is defined as follows:
\abovedisplayskip=5pt
\belowdisplayskip=5pt
\begin{align}
    & \mathbf{H}_c = \text{Conv}(\mathbf{H}_v), \\
    & \mathbf{P}_c = \text{Alternate}(\mathbf{H}_v,\mathbf{H}_c).
\end{align}
Here, $\text{Conv}()$ is the convolutions block (e.g. FSMN~\cite{zhang2018deep}) for modeling local dependencies, $\mathbf{H}_c = \{\mathbf{h}_1^c,\mathbf{h}_2^c,...,\mathbf{h}_{n/2}^c\}$.
$\text{Alternate()}$ is a function to concatenate the visual representations (the blue one) and prediction representations (the red one) alternately (as illustrated in Fig \ref{upsample}), $\mathbf{P}_c = \{\mathbf{h}_1^v,\mathbf{h}_1^c,\mathbf{h}_2^v,\mathbf{h}_2^c,...,\mathbf{h}_{n/2}^v,\mathbf{h}_{n/2}^c\}$. 

Optical flow has been widely used for action recognition in videos~\cite{simonyan2014two}, which utilizes the action motion across the frames to make the recognition easier, as the network does not need to estimate motion implicitly.
Inspired by the optical flow, we propose a residual-based prediction module to utilize the video sequence before downsampling for residual calculating, which obtains the action motion by calculating the difference between adjacent frames.
The residual-based prediction is represented as follows:
\begin{align}
    & \mathbf{H}_d  = \text{Diff}(\mathbf{X}_v), \\
    & \mathbf{H}_r = \mathbf{H}_v + \text{FFN}(\mathbf{H}_d), \\
    & \mathbf{P}_r = \text{Alternate}(\mathbf{H}_v,\mathbf{H}_r).
\end{align}
Here, $\text{Diff}()$ is a residual calculating function, $\mathbf{H}_d = [\mathbf{x}_2^v- \mathbf{x}_1^v,\mathbf{x}_4^v- \mathbf{x}_3^v,...,\mathbf{x}_n^v- \mathbf{x}_{n-1}^v]$. $\text{FFN}()$ means a feed-forward module and $\text{Alternate}()$ is a function to concatenate representations alternately. 

Finally, we generate the final prediction representations by concatenating the $\mathbf{P}_c$ and $\mathbf{P}_r$, which efficiently learns complementarity between contextual and residual information.
If a higher factor $d$ of down-up sampling is required, we can incorporate multi-layer prediction modules.
What's more, by modifying the stride parameter of the convolution in the contextual-based prediction module, we can control the number of prediction representations.
So, with the help of the multi-layer prediction modules and stride modification, we can achieve arbitrary factor $d$ of down-up sampling.
This provides flexibility in adapting model to different scenarios and requirements.

\vspace{-9pt}
\subsection{Visual-Audio Alignment Loss}
\vspace{-3pt}
Compared with the video-only upsampling, the upsampling of AVSR has the advantage of utilizing audio representations for modeling alignment relationships.
Thus, we propose an additional loss function, called visual-audio alignment (VA-Align) loss, to improve the performance of upsampling.
The VA-Align loss is applied to the cross-attention after the concatenating (as shown in Fig \ref{upsample} lower part), which leverages the constraint boundaries of the visual representations to guide the cross-attention learning towards the correct frames.
The formulation of the VA-Align loss is as follows:
%The VA-Align loss is applied to the cross-attention after the concatenating of the two prediction representations (as shown in Fig \ref{upsample} lower part) and leverages the constraint boundaries of visual representations to guide the cross-attention learning towards the correct frames, which is formulated as:
%So, to improve the performance of upsampling, we introduce an additional loss function, called the visual-audio alignment (VA-Align) loss, which is applied to the cross-attention after the concatenating of the two prediction representations (as show in Fig \ref{upsample}).
%The VA-align loss leverages the constraint boundaries of visual representations to guide the cross-attention to focus on the correct time frames, which is formulated as:
\begin{equation}
\mathcal{L}_{\text {VA-Align}}=
\left | n - \sum\limits_{i=1}^{n} \sum\limits_{j=-l}^{r}  \alpha_{i,i+j}
\right |,
\end{equation}
where $\alpha_{i,j} \in (0,1)$ represents the attention score of $i$-th video frame and $j$-th audio frame, and $n$ is the length of the video sequence.
%$l$ and $r$ is the number of frames that looks back to the past and looks ahead to the future at each time step.
$l$ and $r$ is the number of the past and future attention score to be summed at each time step, which aims to pay more attention on local and correct time.
Note that if multi-layer prediction modules are adopted, each layer will have a VA-Align loss for assisting upsampling.

\begin{figure}[!htb]
	\centering
        \vspace{-7pt}
	\includegraphics[scale=0.3]{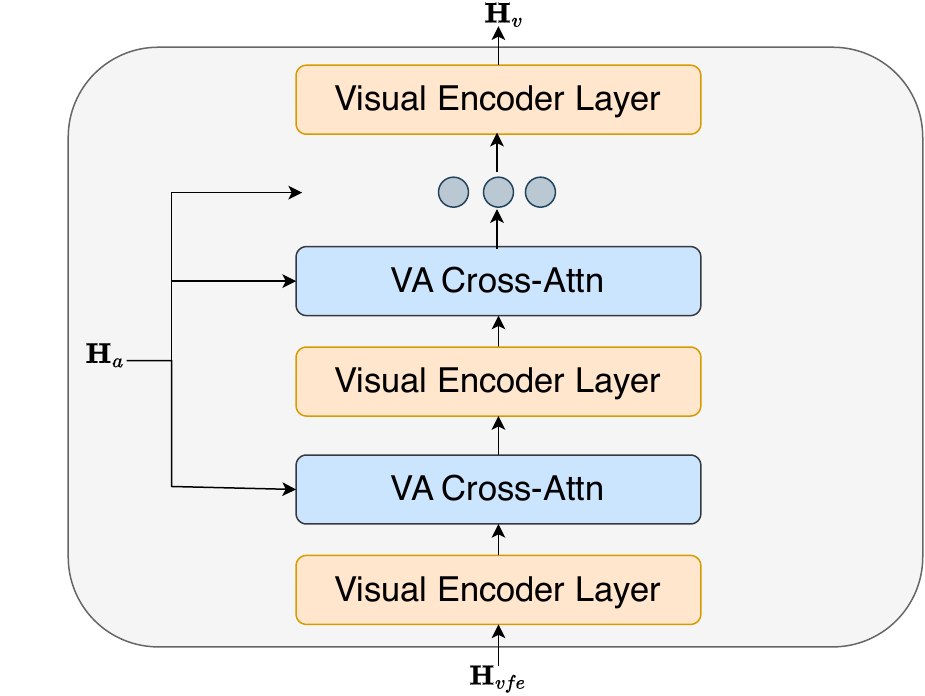}
        \vspace{-7pt}
	\caption{The architecture of cross-layer attention fusion.
	}
	\label{cross-layer}
        \vspace{-20pt}
\end{figure}

\subsection{Cross-Layer Attention Fusion}
\vspace{-3pt}
Fig \ref{cross-layer} depicts the architecture of the visual encoder with cross-layer attention fusion.
Notably, we introduce a visual-audio cross-attention after each visual encoder layer, which takes the visual hidden representations as the query vector and the audio encoder output as the key and value vectors.
In contrast to previous approaches that simply concatenate the modalities together, our proposed method achieves a more effective fusion of different modalities by capturing the modality dependencies within each encoder layer.

\vspace{-4pt}
\subsection{Training Strategy}
\label{strategy}
\vspace{-3pt}
As shown in Fig \ref{model_all}, we adopt a fusion module for audio and visual modalities, consisting of a visual-audio (VA) cross-attention and an audio-visual (AV) cross-attention.
Contrary to the VA cross-attention, the AV cross-attention takes the audio encoder output as the query vector and the visual encoder output as the key and value vectors.
It's worth noting that for the concatenation of the two cross-attention outputs, they must have the same length.
Hence, it becomes necessary to upsample the video sequence in order to match the length of audio sequence.
If we only adopt AV cross-attention without upsampling, the downsampling will seriously degrade the recognition performance, which is investigated in following experiments.
There are three loss functions defined, namely the VA-Align, CTC and CE losses. The types are jointly trained, as follows:
\begin{equation}\label{eq:CER-COMPUTE}
\begin{aligned}
\mathcal L = (1-\lambda_{1}-\lambda_{2})\mathcal L_{\text{CE}} + {\lambda}_{1} \mathcal L_{\text{VA-Align}} +  {\lambda}_{2} \mathcal L_{\text{CTC}},
\end{aligned}
\end{equation}
where ${\lambda}_{1}$ and ${\lambda}_{2}$ are interpolation factors, which are set to $0.2$ and $0.3$ respectively.

\vspace{-10pt}
\section{EXPERIMENTS}
\vspace{-6pt}
\subsection{Experimental Setup}
\vspace{-3pt}
 
We use MISP-AVSR corpus~\cite{chen2022audio,chen2022first}, an audio-visual Mandarin conversational dataset, to evaluate our proposed Hourglass-AVSR model.
This corpus contains 106.09 hours audio and video data for training collected by far/middle/near microphones (106.09 * 3 = 318.27 hours audio data) and far/middle cameras in 34 real-home TV rooms (106.09 * 2 = 212.18 hours video data).
And we report concatenated minimum permutation character error rate (cpCER) on four audio-visual test sets from MISP 2021/2022 challenges, namely eval21 (13.32 hours far-field data), eval21-mid (13.32 hours middle-field data), eval22 (3.09 hours far-field data) and dev22 (3.13 hours far-field data).
We use oracle label and speaker diarization (SD) results to segment test sets, respectively.
Particularly, the SD model achieves 9.4\% diarization error rate (DER) on dev22 set.
The audio data is pre-processed by the WPE and GSS algorithms to extract enhanced clean 1-channel signals for each speaker.
The video data is prepared to extract $112\times112\times3$ RGB lip tracks.

In ASR and AVSR models, the audio/visual encoder consists of 24/12-layer Branchformer~\cite{peng2022branchformer}, and the decoder contains 6 Transformer layers~\cite{DBLP:booktitles/corr/VaswaniSPUJGKP17}.
The dimension of the attention and feed-forward layer is 256 with 4 heads and 2048, respectively.
The visual front-end is an 8-layer ResNet3D module, whose channels range from 16 to 64 and kernel size is 3.
The frame number of VA-Align loss that looks back to the past ($l$) and looks ahead to the future ($r$) sets to 6.
All models are trained on four 32GB memory V100 RTX GPUs with a maximum trainable epoch of 75.
We initialize the AVSR model with the ASR model, similar to other works in MISP challenges.
%Note that our source code and models will be \textbf{opened} in the near future.

\vspace{-3pt}
\subsection{Comparison to Improvement of Different AVSR Models }
\vspace{-2pt}
To better demonstrate the advantages of our proposed Hourglass-AVSR model, we conduct a detailed comparison to the performance improvements of the AVSR models in the MISP2021 and MISP2022 challenges, such as multi-channel AVSR (MC-AVSR)~\cite{xu2022channel}, conformer-based AVSR (CF-AVSR)~\cite{wangxiaomi}, front-end joint AVSR (FE-AVSR)~\cite{wang2022sjtu}, cross-attention fusion AVSR (CA-ASR)~\cite{guo2023npu} and multi-level fusion AVSR (ML-AVSR)~\cite{li2023xmu}.
%Because the ASR architecture design and data augmentation is not our major focus, we do not further elaborate and 
As shown in Table~\ref{table1}, our proposed Hourglass-AVSR model outperforms ASR model, leading to 12.9\% average relative cpCER reduction on test sets.
Compared with other SOTA AVSR models, our Hourglass-AVSR model shows the best average relative improvement.
The MC-AVSR, CF-AVSR and FE-AVSR models all employ the pre-computed visual feature from a fine-tuned lipreading model, achieving 1.5\% or 1.4\% relative cpCER reduction on dev21 set.
We believe that visual frontend module need to be trained jointly within the AVSR model for learning fine-grained visual representations.
So, it's necessary to reduce the computation costs and achieve a computational efficiency model.
Besides, cross-layer attention fusion (CLAF) plays a significant role in our model, decreasing the cpCER from 27.6\%/26.0\%/29.7\% to 26.1\%/24.5\%/28.2\% on dev21, dev22, dev22$^{sd}$ sets.

\begin{table}[!ht]
\vspace{-12pt}
\caption{Results for various ASR and AVSR models on dev21, dev22 and dev22$^{sd}$ sets (cpCER \%). The average relative improvement (\%) between ASR and AVSR can be found in last column. }
\centering
%\begin{threeparttable}[t]
\vspace{3pt}

\setlength{\tabcolsep}{3pt}
\label{table1}
\begin{threeparttable}

\scalebox{0.87}{
\begin{tabular}{lccccccc}
\toprule
\hline
\multicolumn{1}{c}{\multirow{2}{*}{Model}} & \multicolumn{3}{c}{ASR} & \multicolumn{3}{c}{AVSR} & \multirow{2}{*}{Imprv.} \\ \cmidrule(r){2-4} \cmidrule(r){5-7}
\multicolumn{1}{c}{}        & dev21 & dev22 & dev22$^{sd}$ & dev21  & dev22 & dev22$^{sd}$ &                                   \\ \hline
MC-AVSR\tnote{*}~\cite{xu2022channel}                   & 32.5  & -     & -       & 32.0   & -     & -       & 1.5                               \\
CF-AVSR\tnote{*}~\cite{wangxiaomi}                   & 41.9  & -     & -       & 41.3   & -     & -       & 1.4                               \\
FE-AVSR~\cite{wang2022sjtu}                   & 39.8  & -     & -       & 39.3   & -     & -       & 1.5                               \\
CA-AVSR~\cite{guo2023npu}                 & -     & \textbf{26.6}  & \textbf{30.7}    & -      & 25.7  & 29.7    & 3.4                               \\
ML-AVSR~\cite{li2023xmu}                   & -     & -     & 32.7    & -      & -     & 29.4    & 10.1                              \\ \hline
Hourglass-AVSR                  & \textbf{30.0}  & 28.2  & 32.3    & \textbf{26.1}   & \textbf{24.5}  & \textbf{28.2}    & \textbf{12.9}                              \\ 
\quad -CLAF                  & 30.0  & 28.2  & 32.3    & 27.6   & 26.0  & 29.7    & 7.9                              \\ \hline
\bottomrule
\end{tabular}
}

\begin{tablenotes}
    \footnotesize
		\item *: The ablation experimental results of integrating visual modules.
\end{tablenotes}
\end{threeparttable}

\vspace{-18pt}
\end{table}
\vspace{-3pt}
\subsection{Impact of the Different Down-Up Sampling Approaches}
\vspace{-2pt}
We further explore the impact of different down-up sampling approaches and conduct ablation experiments on our proposed model, as shown in Table~\ref{table2}.
Although AVSR model achieves remarkable recognition performance that leads to 12.9\% and 20.8\% relative cpCER reduction on far-field and middle-field test sets respectively, the visual module of AVSR brings larger computation costs compared with ASR model, increasing the GFlops from 32.9 to 177.1.
When incorporating the 2/3/4$\times$ downsampling into AVSR model without upsampling, which only uses AV cross-attention for modality fusion introduced in Section \ref{strategy}, the recognition performance of the model is weakened seriously.
To compensate for the performance loss caused by the downsampling, we use contextual-based upsampling to capture more context information, and the performance can be significantly improved, achieving 1.8\%/1.9\% (27.7\%/25.9\% $\to$ 27.2\%/25.4\%) relative cpCER reduction on dev21 and dev22 sets of 4$\times$ downsampling.
With the help of residual-based prediction, the upsampling can achieve better prediction performance and further reduce the cpCER compared to using contextual-based prediction alone. 
Meanwhile, VA-Align loss captures inherent alignments between audio and video and achieves cpCERs of 26.6\%, 31.8\%, 24.9\% and 28.8\% on four test sets of 4$\times$ downsampling.
%Meanwhile, VA-Align loss also improves the performance and achieves cpCERs of 26.6\%, 31.8\%, 24.9\% and 28.8\% on four test sets of 4$\times$ downsampling.
%In conclusion, the results demonstrate that our proposed upsampling approaches successfully compensate 68.8\% and 50\% performance loss of 2$\times$ and 4$\times$ downsampling with low computation cost, especially on the 4$\times$ downsampling, leading to 54.2\% relative GFlops reduction.
In conclusion, the results demonstrate that our proposed upsampling approaches achieve a comparable cpCER with lower computation costs, especially on the 4$\times$ downsampling, leading to 54.2\% (177.1 $\to$ 81.1) relative GFlops reduction.

\begin{table}[!ht]
\vspace{-7pt}
\caption{Results for various down-up sampling approaches on four test sets (cpCER \%). GFlops stands for the number of floating-point operations that a model performs per second input (100 frames audio and 25 frames video), which is a measure of model efficiency.}
%\caption{Results for various down-up sampling approaches on four test sets (cpCER \%). Giga floating point operations per second (GFlops) is a measure of model efficiency.}
\vspace{3pt}
\centering
%\begin{threeparttable}[t]

\setlength{\tabcolsep}{3pt}
\label{table2}
\scalebox{0.95}{
\begin{tabular}{lccccc}
\toprule
\hline
\multicolumn{1}{c}{Model} & GFlops & dev21 & dev21-mid & dev22 & dev22$^{sd}$ \\ \hline
ASR                       & 32.9   & 30.0  & 39.0         & 28.2  & 32.3    \\
Hourglass-AVSR                  & 177.1  & \textbf{26.1}  & \textbf{30.9}         & \textbf{24.5}  & \textbf{28.2}    \\ \hline
\quad+2$\times$ Downsampling          & 110.1  & 26.7  & 31.9         & 25.0  & 28.8    \\
\quad \quad+Context                  & 112.8  & 26.5  & 31.6         & 24.8  & 28.6    \\
\quad \quad \quad+Residual                  & 116.3  & 26.4  & 31.4         & 24.7  & 28.6    \\
\quad \quad \quad \quad+VA-Align                 & 116.7  & \textbf{26.4}  & \textbf{31.3}         & \textbf{24.7}  & \textbf{28.5}    \\ \hline
\quad +3$\times$ Downsampling          & 88.4   & 27.2  & 32.8         & 25.5  & 29.4    \\ \hline
\quad +4$\times$ Downsampling          & 77.6   & 27.7  & 33.6         & 25.9  & 29.9    \\
\quad \quad+Context                 & 79.0   & 27.2  & 32.7         & 25.4  & 29.4    \\
\quad \quad \quad+Residual                  & 80.9   & 26.8  & 32.1         & 25.1  & 29.0    \\
\quad \quad \quad \quad+VA-Align                 & 81.1   & \textbf{26.6}  & \textbf{31.8}         & \textbf{24.9}  & \textbf{28.8}    \\ \hline
\bottomrule
\end{tabular}
}
\vspace{-18pt}
\end{table}

\subsection{Comparison with MISP2021 and MISP2022 top models}
\vspace{-2pt}

As shown in Table~\ref{table3}, we compare the results of our proposed Hourglass-AVSR model with the top-ranking models in MISP2021 and MISP2022 challenges.
It can be observed that our Hourglass-AVSR model trained without data augmentation and simulation still obtains an outstanding results, even surpassing the results of CF-AVSR and CA-AVSR models (2$^{nd}$ in MISP 2021/2022) by achieving 26.1\%, 28.2\% and 32.0\% cpCER on dev21, dev22$^{sd}$ and eval22$^{sd}$ sets.
%n both challengs by 5.1\% (27.5\% $\to$ 26.1\%) and 5.0\% (33.7\% $\to$ 32.0\%) relative cpCER improvement on dev21 and eval22$^{sd}$ sets.
The ranking 1$^{st}$ MC-AVSR and multi-channel conformer-based AVSR (MCC-AVSR)~\cite{xu2023nio} models achieve the lowest results by large-scale data simulation and strong multi-channel based audio encoder with large computational overhead.
we will continue to validate our Hourglass-AVSR model on large-scale datasets and different acoustic model in the near future.

\begin{table}[!ht]
\vspace{-15pt}
\caption{Results of top-ranking models in MISP2021/2022 challenges and ours proposed model on dev21, dev22$^{sd}$ and eval22$^{sd}$ sets (cpCER \%).}
\centering
%\begin{threeparttable}[t]
\vspace{3pt}

\setlength{\tabcolsep}{3pt}
\label{table3}
\scalebox{0.95}{
\begin{tabular}{cccccc}
\toprule
\hline
Model    & MISP Rank & Data(hrs) & dev21 & dev22$^{sd}$ & eval22$^{sd}$ \\ \hline
MC-AVSR~\cite{xu2022channel}   & 2021 1$^{st}$  & 3,498      & \textbf{25.1}  & -       & -        \\
CF-AVSR~\cite{wangxiaomi}   & 2021 2$^{nd}$  & 3,000      & 27.5  & -       & -        \\
FE-AVSR~\cite{wang2022sjtu}   & 2021 3$^{rd}$  & \textbf{318}       & 39.3  & -       & -        \\
MCC-AVSR~\cite{xu2023nio}  & 2022 1$^{st}$  & 3,498      & -     & -       & \textbf{29.6}     \\
CA-AVSR~\cite{guo2023npu}  & 2022 2$^{nd}$  & 1,272      & -     & 29.7    & 33.7     \\
ML-AVSR~\cite{li2023xmu}   & 2022 3$^{rd}$  & 2,000      & -     & 29.4    & 33.9     \\ \hline
Hourglass-AVSR & -         & \textbf{318}       & 26.1  & \textbf{28.2}    & 32.0     \\ \hline
\bottomrule
\end{tabular}
}
\vspace{-22pt}
\end{table}

\section{Conclusions}
\vspace{-5pt}
In this work, we propose a down-up sampling-based computational efficiency model for AVSR (Hourglass-AVSR).
%To enjoys both high efficiency and high performance, we propose a context and residual aware video upsampling approach to compensate for performance loss caused by the downsampling.
To enjoys high efficiency and performance, we propose a context and residual aware video down-up sampling approach.
Besides, we also propose a visual-audio alignment loss to learn boundary constraints explicitly and a cross-layer attention fusion to capture modality dependencies.
%inherent alignments.
%Finally, in order to capture the modality dependencies within each encoder layer, we further propose a cross-layer attention fusion approach.
Evaluated on the real audio-visual conversational corpus MISP-AVSR, our proposed Hourglass-AVSR model outperforms ASR model by 12.9\% and 20.8\% relative cpCER reduction on far-field and middle-field sets, which achieves the best average relative improvement compared with other SOTA AVSR models.
%Compared with other SOTA AVSR models, our model achieves the best average relative improvement of visual module on several test sets.
%Moreover, our proposed upsampling approaches successfully compensate 68.8\% performance loss of downsampling with only 1/2 computation cost.
Moreover, on the benefit of our down-up sampling approach, Hourglass-AVSR model reduces 54.2\% overall computation costs with only few performance degradation.
In the future, we will investigate an industrial-level ASR model trained on large-scale data into our proposed Hourglass-AVSR model for real-world applications.


\begin{spacing}{0.1}
\bibliographystyle{IEEEbib}
\bibliography{strings,refs}

\begin{thebibliography}{10}

\bibitem{li2021recent}
Jinyu Li et~al.,
\newblock ``Recent advances in end-to-end automatic speech recognition,''
\newblock {\em Proc. APSIPA}, vol. 11, no. 1, 2022.

\bibitem{Yu2022M2MeT}
Fan Yu, Shiliang Zhang, Yihui Fu, Lei Xie, Siqi Zheng, Zhihao Du, et~al.,
\newblock ``M2{M}e{T}: The {ICASSP} 2022 multi-channel multi-party meeting
  transcription challenge,''
\newblock in {\em Proc. ICASSP}. IEEE, 2022, pp. 6167--6171.

\bibitem{Yu2022Summary}
Fan Yu, Shiliang Zhang, Pengcheng Guo, Yihui Fu, Zhihao Du, Siqi Zheng, Lei
  Xie, et~al.,
\newblock ``Summary on the {ICASSP} 2022 multi-channel multi-party meeting
  transcription grand challenge,''
\newblock in {\em Proc. ICASSP}. IEEE, 2022, pp. 9156--9160.

\bibitem{fiscus2007rich}
Jonathan~G Fiscus, Jerome Ajot, and John~S Garofolo,
\newblock ``The rich transcription 2007 meeting recognition evaluation,''
\newblock in {\em Proc.MTPH}, pp. 373--389. Springer, 2007.

\bibitem{fiscus2006rich}
Jonathan~G Fiscus, Jerome Ajot, Martial Michel, et~al.,
\newblock ``The rich transcription 2006 spring meeting recognition
  evaluation,''
\newblock in {\em Proc. MLMI}. Springer, 2006, pp. 309--322.

\bibitem{chen2022audio}
Hang Chen, Jun Du, Yusheng Dai, Chin~Hui Lee, Sabato~Marco Siniscalchi, Shinji
  Watanabe, Odette Scharenborg, Jingdong Chen, et~al.,
\newblock ``{Audio-visual speech recognition in misp2021 challenge: Dataset
  release and deep analysis},''
\newblock in {\em Proc. INTERSPEECH}. ISCA, 2022, pp. 1766--1770.

\bibitem{tamura2015audio}
Satoshi Tamura, Hiroshi Ninomiya, Norihide Kitaoka, Shin Osuga, Yurie Iribe,
  Kazuya Takeda, and Satoru Hayamizu,
\newblock ``Audio-visual speech recognition using deep bottleneck features and
  high-performance lipreading,''
\newblock in {\em Proc. APSIPA}. IEEE, 2015, pp. 575--582.

\bibitem{petridis2018end}
Stavros Petridis, Themos Stafylakis, Pingehuan Ma, Feipeng Cai, Georgios
  Tzimiropoulos, and Maja Pantic,
\newblock ``End-to-end audiovisual speech recognition,''
\newblock in {\em Proc. ICASSP}. IEEE, 2018, pp. 6548--6552.

\bibitem{makino2019recurrent}
Takaki Makino, Hank Liao, Yannis Assael, Brendan Shillingford, Basilio Garcia,
  Otavio Braga, and Olivier Siohan,
\newblock ``Recurrent neural network transducer for audio-visual speech
  recognition,''
\newblock in {\em Proc. ASRU}. IEEE, 2019, pp. 905--912.

\bibitem{xu2020discriminative}
Bo~Xu, Cheng Lu, Yandong Guo, and Jacob Wang,
\newblock ``Discriminative multi-modality speech recognition,''
\newblock in {\em Proc. CVPR}. IEEE/CVF, 2020, pp. 14433--14442.

\bibitem{son2017lip}
Joon Son~Chung, Andrew Senior, Oriol Vinyals, and Andrew Zisserman,
\newblock ``Lip reading sentences in the wild,''
\newblock in {\em Proc. CVPR}. IEEE/CVF, 2017, pp. 6447--6456.

\bibitem{deep2022afouras}
Triantafyllos Afouras, Joon~Son Chung, Andrew Senior, Oriol Vinyals, and Andrew
  Zisserman,
\newblock ``Deep audio-visual speech recognition,''
\newblock in {\em Proc. TPAMI}. IEEE, 2022, pp. 8717--8727.

\bibitem{ma2021end}
Pingchuan Ma, Stavros Petridis, and Maja Pantic,
\newblock ``{End-to-end audio-visual speech recognition with conformers},''
\newblock in {\em Proc. ICASSP}. IEEE, 2021, pp. 7613--7617.

\bibitem{xu2023nio}
Gaopeng Xu, Xianliang Wang, Sang Wang, Junfeng Yuan, Wei Guo, Wei Li, and Jie
  Gao,
\newblock ``{The NIO System for audio-visual diarization and recognition in
  MISP challenge 2022},''
\newblock in {\em Proc. ICASSP}. IEEE, 2023, pp. 1--2.

\bibitem{xu2022channel}
Gaopeng Xu, Song Yang, Wei Li, Song Wang, Guo Wei, Junfeng Yuan, and Jie Gao,
\newblock ``Channel-wise av-fusion attention for multi-channel audio-visual
  speech recognition,''
\newblock in {\em Proc. ICASSP}. IEEE, 2022, pp. 9251--9255.

\bibitem{sterpu2018attention}
George Sterpu, Christian Saam, and Naomi Harte,
\newblock ``{Attention-based audio-visual fusion for robust automatic speech
  recognition},''
\newblock in {\em Proc. MI}. ACM, 2018, pp. 111--115.

\bibitem{wei2020attentive}
Liangfa Wei, Jie Zhang, Junfeng Hou, and Lirong Dai,
\newblock ``{Attentive fusion enhanced audio-visual encoding for Transformer
  based robust speech recognition},''
\newblock in {\em Proc. APSIPA ASC}. IEEE, 2020, pp. 638--643.

\bibitem{wu2021audio}
Yifei Wu, Chenda Li, Song Yang, Zhongqin Wu, and Yanmin Qian,
\newblock ``{Audio-visual multi-Talker speech recognition in a cocktail
  party},''
\newblock in {\em Proc. INTERSPEECH}. ISCA, 2021, pp. 3021--3025.

\bibitem{li2023xmu}
Tao Li, Haodong Zhou, Jie Wang, Qingyang Hong, and Lin Li,
\newblock ``{The XMU system for audio-visual diarization and recognition in
  MISP challenge 2022},''
\newblock in {\em Proc. ICASSP}. IEEE, 2023, pp. 1--2.

\bibitem{he2016deep}
Kaiming He, Xiangyu Zhang, Shaoqing Ren, and Jian Sun,
\newblock ``{Deep residual learning for image recognition},''
\newblock in {\em Proc. CVPR}. IEEE/CVF, 2016, pp. 770--778.

\bibitem{simonyan2014very}
Karen Simonyan and Andrew Zisserman,
\newblock ``Very deep convolutional networks for large-scale image
  recognition,''
\newblock in {\em Proc. ICLR}, 2015, pp. 1--14.

\bibitem{tran2015learning}
Du~Tran, Lubomir Bourdev, Rob Fergus, Lorenzo Torresani, and Manohar Paluri,
\newblock ``Learning spatiotemporal features with 3d convolutional networks,''
\newblock in {\em Proc. CVPR}. IEEE/CVF, 2015, pp. 4489--4497.

\bibitem{lee2018motion}
Myunggi Lee, Seungeui Lee, Sungjoon Son, Gyutae Park, and Nojun Kwak,
\newblock ``Motion feature network: Fixed motion filter for action
  recognition,''
\newblock in {\em Proc. ECCV}, 2018, pp. 387--403.

\bibitem{lin2019tsm}
Ji~Lin, Chuang Gan, and Song Han,
\newblock ``Tsm: Temporal shift module for efficient video understanding,''
\newblock in {\em Proc. CVPR}. IEEE/CVF, 2019, pp. 7083--7093.

\bibitem{tran2018closer}
Du~Tran, Heng Wang, Lorenzo Torresani, Jamie Ray, Yann LeCun, and Manohar
  Paluri,
\newblock ``A closer look at spatiotemporal convolutions for action
  recognition,''
\newblock in {\em Proc. CVPR}. IEEE/CVF, 2018, pp. 6450--6459.

\bibitem{xie2018rethinking}
Saining Xie, Chen Sun, Jonathan Huang, Zhuowen Tu, and Kevin Murphy,
\newblock ``Rethinking spatiotemporal feature learning: Speed-accuracy
  trade-offs in video classification,''
\newblock in {\em Proc. ECCV}, 2018, pp. 305--321.

\bibitem{shou2017cdc}
Zheng Shou, Jonathan Chan, Alireza Zareian, Kazuyuki Miyazawa, and Shih-Fu
  Chang,
\newblock ``{CDC}: Convolutional-de-convolutional networks for precise temporal
  action localization in untrimmed videos,''
\newblock in {\em Proc. CVPR}. IEEE/CVF, 2017, pp. 5734--5743.

\bibitem{peng2022branchformer}
Yifan Peng, Siddharth Dalmia, Ian Lane, and Shinji Watanabe,
\newblock ``{Branchformer: Parallel MLP-attention architectures to capture
  local and global context for speech recognition and understanding},''
\newblock in {\em Proc. ICML}. PMLR, 2022, pp. 17627--17643.

\bibitem{zhang2018deep}
Shiliang Zhang, Ming Lei, Zhijie Yan, and Lirong Dai,
\newblock ``Deep-fsmn for large vocabulary continuous speech recognition,''
\newblock in {\em Proc. ICASSP}. IEEE, 2018, pp. 5869--5873.

\bibitem{simonyan2014two}
Karen Simonyan and Andrew Zisserman,
\newblock ``Two-stream convolutional networks for action recognition in
  videos,''
\newblock {\em Proc. NIPS}, vol. 27, 2014.

\bibitem{chen2022first}
Hang Chen, Hengshun Zhou, Jun Du, Chin-Hui Lee, Jingdong Chen, Shinji Watanabe,
  et~al.,
\newblock ``{The first multimodal information based speech processing (MISP)
  challenge: Data, tasks, baselines and results},''
\newblock in {\em Proc. ICASSP}. IEEE, 2022, pp. 9266--9270.

\bibitem{DBLP:booktitles/corr/VaswaniSPUJGKP17}
Ashish Vaswani, Noam Shazeer, Niki Parmar, Jakob Uszkoreit, Llion Jones,
  Aidan~N Gomez, {\L}ukasz Kaiser, and Illia Polosukhin,
\newblock ``Attention is all you need,''
\newblock in {\em Proc. NeurIPS}, 2017, pp. 5998--6008.

\bibitem{wangxiaomi}
Quandong Wang, Xinyu Cai, Weiji Zhuang, Yuxiang Kong, Yongqing Wang, Junnan Wu,
  et~al.,
\newblock ``The xiaomi-talkfreely system for audio-visual speech recognition in
  misp challenge 2021,''
\newblock 2022.

\bibitem{wang2022sjtu}
Wei Wang, Xun Gong, Yifei Wu, Zhikai Zhou, Chenda Li, Wangyou Zhang, Bing Han,
  and Yanmin Qian,
\newblock ``The sjtu system for multimodal information based speech processing
  challenge 2021,''
\newblock in {\em Proc. ICASSP}. IEEE, 2022, pp. 9261--9265.

\bibitem{guo2023npu}
Pengcheng Guo, He~Wang, Bingshen Mu, Ao~Zhang, and Peikun Chen,
\newblock ``{The NPU-ASLP system for audio-visual speech recognition in MISP
  2022 challenge},''
\newblock in {\em Proc. ICASSP}. IEEE, 2023, pp. 1--2.

\end{thebibliography}
\end{spacing}

\end{document}